\documentclass{article}
\usepackage{amssymb,amsmath,dsfont,alife10}
\usepackage[dvips]{psfrag}

\newcommand{\beq}{\begin{eqnarray}}
\newcommand{\eeq}{\end{eqnarray}}
\newcommand{\beqn}{\begin{eqnarray*}}
\newcommand{\eeqn}{\end{eqnarray*}}

\newcommand{\eM}     {$\epsilon$-machine}
\newcommand{\eMs}    {$\epsilon$-machines}
\newcommand{\EM}     {$\epsilon$-Machine}
\newcommand{\EMs}    {$\epsilon$-Machines}

\newcommand{\BiInfinity}        { \stackrel{\leftrightarrow} {S} }

\newcommand{\Past}              { \stackrel{\leftarrow} {S} }
\newcommand{\past}              { {\stackrel{\leftarrow} {s}} }
\newcommand{\pastprime}         { {\past}^{\prime}}
\newcommand{\Future}            { \stackrel{\rightarrow}{S} }

\newcommand{\AllPasts}          { { \stackrel{\leftarrow} {\rm {\bf S}} } }

\newcommand{\CausalState}       { {\cal S} }

\newcommand{\CausalStateSet}    { \boldsymbol{\CausalState} }
\newcommand{\TransProbSet}      { \boldsymbol{\cal T} }

\newcommand{\NextObservable}    { {\stackrel{\rightarrow} {S}}^1 }

\newcommand{\Prob}              { {\rm P} }

\newcommand{\Cmu}               { C_\mu }

\newcommand{\Alphabet}           { \mathcal{A} }

\begin{document}

\title{Hierarchical Self-Organization in the Finitary Process Soup}
\author{Olof G\"{o}rnerup$^{1, 3,\dagger}$ and
James P.  Crutchfield$^{2, 3,*}$\\
$^1$ Department of Physical Resource Theory, Chalmers University of Technology, 412 96 G\"{o}teborg, Sweden\\
$^2$ Center for Computational Science \& Engineering and Physics Department,\\
University of California, Davis, One Shields Avenue, Davis CA 95616\\
$^3$ Santa Fe Institute, 1399 Hyde Park Road, Santa Fe, NM 87501\\
$^\dagger$olof.gornerup@fy.chalmers.se\\
$^*$chaos@cse.ucdavis.edu
}

\maketitle
 

\begin{abstract}
Current analyses of genomes from numerous species show that the diversity of
organism's functional and behavioral characters is not proportional to the
number of genes that encode the organism. We investigate the hypothesis that
the diversity of organismal character is due to hierarchical organization.
We do this with the recently introduced model of the
\emph{finitary process soup}, which allows for a detailed
mathematical and quantitative analysis of the population dynamics of structural
complexity. Here we show that global complexity in the finitary process soup
is due to the emergence of successively higher levels of organization, that the
hierarchical structure appears spontaneously, and that the process of
structural innovation is facilitated by the discovery and maintenance of
relatively noncomplex, but general individuals in a population.
\end{abstract}


\section{Introduction}

Recent estimates have shown that the genomes of many species consist of a
surprisingly similar number of genes despite some being markedly more
sophisticated and diverse in their behaviors. Humans have only 30\% more
genes that the worm \emph{Caenorhabditis elegans}; humans, mice, and rats
have nearly the same number \cite{Lync03a,RGSPC04a}. Moreover, many of
those genes serve to maintain elementary processes and are shared across
species, which greatly reduces the number of genes available to account for
diversity. One concludes that individual genes cannot directly code for the
full array of individual functional and morphological characters of a species,
as genetic determinism would have it. From what, then, do the sophistication
and diversity of organismal form and behavior arise?

Here we investigate the hypothesis that these arise from a hierarchy of
interactions between genes and between interacting gene complexes. A hierarchy
of gene interactions, being comprised of subsets of available genes, allows
for an exponentially larger range of functions and behaviors than direct
gene-to-function coding. We will use a recently introduced pre-biotic
evolutionary model---the \emph{finitary process soup}---of the population
dynamics of structural complexity \cite{Crut04a}. Specifically, we will show
that global complexity in the finitary process soup is due to the emergence
of successively higher levels of organization. Importantly, hierarchical
structure appears spontaneously and is facilitated by the discovery and
maintenance of
relatively noncomplex, but general individuals in a population. These results,
in concert with the minimal assumptions and simplicity of the finitary process
soup, strongly suggest that an evolving system's sophistication, complexity,
and functional diversity derive from its hierarchical organization. 


\section{Modeling Pre-Biology}

Prior to the existence of highly sophisticated entities acted on by
evolutionary forces, replicative objects relied on far more basic
mechanisms for maintenance and growth. However, these objects managed to
transform, not only themselves, but also indirectly the very transformations
by which they changed \cite{Rossler1} in order to eventually support the
mechanisms of natural selection. How did the transition from raw interaction
to evolutionary change take place? Is it possible to pinpoint generic
properties, however basic, that would have enabled a system of simple
interacting objects to take the first few steps towards biotic organization?

To explore these questions in terms of structural complexity we
developed a theoretical model borrowing from computation theory \cite{Hopc79}
and computational mechanics \cite{Crut88a,Crut92c}. In this system---the
\emph{finitary process soup} \cite{Crut04a}---elementary objects,
as represented by \eMs, interact and generate new objects in a well stirred
flow reactor. 

Choosing \eMs\ as the interacting, replicating objects, it turns out, brings
a number of advantages. Most particularly, there is a well developed theory
of their structural properties found in the framework of computational
mechanics. In contrast with individuals in previous, related pre-biotic
models---such as machine language programs
\cite{Rasmussen1,Rasmussen2,Ray91a,Adami1}, tags \cite{Farmer1,Bagley1},
$\lambda$-expressions \cite{Font91a}, and cellular automata
\cite{Crutchfield&Mitchell94a}, \eMs\ have a well defined (and calculable)
notion of structural complexity. For the cases of machine language and
$\lambda$-calculus, in contrast, it is known that algorithms do not even
exist to calculate such properties since these representations are computation
universal \cite{Broo89a}. Another important distinction with prior pre-biotic
models is that the individuals in the finitary process soup do not have two
separate modes of operation---one of representation or storage and one for
functioning and transformation. The individuals are
simply objects whose internal structure determines how they interact. 
The benefit of this when modeling prebiotic evolution is that there is no
assumed distinction between gene and protein \cite{schrodinger-life,Neum66a}
or between data and program \cite{Rasmussen1,Rasmussen2,Ray91a,Adami1}.

\section{\EMs}

Individuals in the finitary process soup are objects that store and
transform information. In the vocabulary of information theory they are
\emph{communication channels} \cite{Cove91a}. Here we focus on a type of
finite-memory channel, called a finitary \eM, as our preferred representation
of an evolving information-processing individual. To understand what this
choice captures we can think of these individuals in terms of how they
compactly describe stochastic processes.

A \emph{process} is a discrete-valued,
discrete-time stationary stochastic information source \cite{Cove91a}. A
process is most directly described by the bi-infinite sequence it produces
of random variables $S_t$ over an alphabet $\Alphabet$:
\begin{equation}
\BiInfinity = ... S_{t-1} S_t S_{t+1} ...
\end{equation}
and the distribution $\Prob(\BiInfinity)$ over those sequences. At each moment
$t$, we think of the bi-infinite sequence as consisting of a history
$\Past_t$ and a future $\Future_t$ subsequence:
$\BiInfinity = \Past_t \Future_t$.

A process stores information in its set $\CausalStateSet$ of
\emph{causal states}. Mathematically, these are the members of the range of
the map $\epsilon: ~\AllPasts \mapsto 2^{\AllPasts}$ from histories to
sets of histories
\begin{equation}
  \epsilon(\past)  = \{ \pastprime | \Prob(\Future|\Past=\past)
  = \Prob(\Future|\Past=\pastprime) \}~,
\end{equation}
where $2^{\AllPasts}$ is the power set of histories $\AllPasts$. That is,
the causal
state $\CausalState$ of a history $\past$ is the \emph{set} of histories that
all have the same probability distribution of futures. The transition from
one causal state $\CausalState_i$ to another $\CausalState_j$ while emitting
the symbol $s \in \Alphabet$ is given by a set of labeled transition matrices:
$\TransProbSet = \{T_{ij}^{(s)}: s \in \Alphabet \}$, in which
\begin{equation}
T_{ij}^{(s)} \equiv \Prob(\CausalState^\prime = \CausalState_j,
  \NextObservable = s| \CausalState = \CausalState_i) ~,
\end{equation}
where $\CausalState$ is the current casual state, $\CausalState^\prime$ its
successor, and $\NextObservable$ the next symbol in the sequence.

A process' \emph{\eM} is the ordered pair $\{\CausalStateSet, \TransProbSet\}$.
Finitary \eMs\ are stochastic finite-state machines with the following
properties \cite{Crut88a}: (i) All recurrent states form a single strongly
connected component. (ii) All transitions are deterministic in the specific
sense that a causal state together with the next symbol determine a unique
next state. (iii) The set of causal states is finite and minimal.

\begin{figure}
\begin{center}
\scalebox{0.55}{\includegraphics{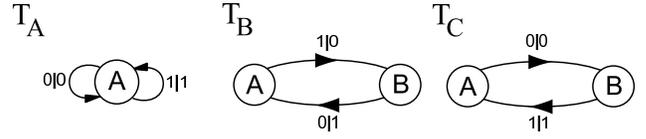}}
\caption{Example \eMs: $T_A$ has a single causal state and, according to its
  transition labels, is the identity function. $T_B$ consists of causal states
  $A$ and $B$ and two transitions. $T_B$ accepts two input strings, either
  $1010\ldots$ or $0101\ldots$, and flips $0$s to $1$s and vice versa as it
  produces an output string. Note that the function's domain and range are
  the same. $T_C$ has the same domain and range as $T_B$, but does not
  exchange $0$s and $1$s.
  }
\label{ExTr}
\end{center}
\end{figure}

In the finitary process soup we use the alphabet
$\Alphabet=\{0|0, 0|1, 1|0, 1|1\}$ consisting of pairs
$in~|~out$ of input and output symbols over a binary alphabet
$\mathcal{B}=\{0, 1\}$. When used in this way \eMs\ read in strings over
$\mathcal{B}$ and emit strings over $\mathcal{B}$. Accordingly, they should be
viewed as mappings from one process $\BiInfinity_{\mathrm{input}}$ to
another $\BiInfinity_{\mathrm{output}}$. They are, in fact, simply functions,
each with a domain (the set of strings that can be read) and with a range
(the set of strings that can be produced). In this way, we consider
\eMs\ as models of objects that store and transform information. In the
following we will take the transitions from each causal state to have equal
probabilities. Figure \ref{ExTr} shows several examples of simple \eMs.

Given that \eMs\ are transformations, one can ask how much processing they
do---how much structure do they add to the inputs when producing an output?
Due to the properties mentioned above, one can answer this question precisely.
Ignoring input and output symbols, the state-to-state transition probabilities
are given by an \eM's \emph{stochastic connection matrix}:
$\mathbf{T}\equiv \sum_{s\in \Alphabet} T^{(s)}$. The causal-state probability
distribution $p_{\CausalStateSet}$ is given by the left eigenvector of
$\mathbf{T}$ associated with eigenvalue $1$ and normalized in probability. If
$M$ is an \eM, then the amount of information storage it has, and can add to
an input process, is given by $M$'s \emph{structural complexity}
\begin{equation}
C_{\mu}(M) \equiv -\sum_{v\in \CausalStateSet}
  p_{\CausalStateSet}(v) \log_2 p_{\CausalStateSet}(v)~.
\end{equation}

\section{\EM\ Interaction}

\eMs\ interact by functional composition. Two machines $T_A$ and $T_B$ that
act on each other result in a third $T_C = T_B \circ T_A$, where $T_C$ (i)
has the domain of $T_A$ and the range of $T_B$ and (ii) is minimized. If
$T_A$ and $T_B$ are incompatible, e.g., the domain of $T_B$ does not overlap
with the range of $T_A$, the interaction produces nothing---it is considered
elastic. During composition the size of the resulting \eM\ can grow very
rapidly (geometrically): $|T_C| \leq |T_B| \times |T_A|$.

\begin{figure}
\begin{center}
\scalebox{0.55}{\includegraphics{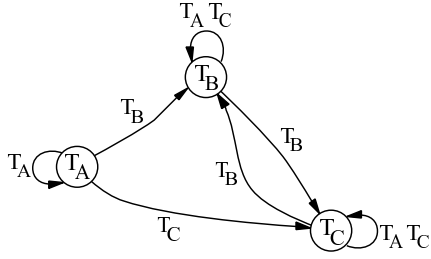}}
\caption{Interaction network for the \eMs\ of Fig. \ref{ExTr}. It
  is a meta-machine.}
\label{ExNet}
\end{center}
\end{figure}

\subsection{Interaction Network}

We monitor the interactions of objects in the soup via the
\emph{interaction network} $\mathcal{G}$. This is represented as a graph
whose nodes correspond to \eMs\ and whose transitions correspond to
interactions. If $T_k=T_j \circ T_i$ occurs in the soup, then the edge
from $T_i$ to $T_k$ is labeled $T_j$. One can represent $\mathcal{G}$
with the binary matrices:
\begin{equation}
\mathcal{G}_{ij}^{(k)}=
\begin{cases}
1 & \text{if $T_k=T_j \circ T_i$} \\
0& \text{otherwise}.
\end{cases}
\label{}
\end{equation}
For the set of \eMs\ in Fig. \ref{ExTr}, for example, we have the
interaction graph shown in Fig. \ref{ExNet} that is given by the matrices:
\begin{center}
$
\mathcal{G}^{(A)} = \left[
	\begin{matrix}
	1 & 0 & 0 \\
	0 & 0 & 0 \\
	0 & 0 & 0 
	\end{matrix}
  \right]~,
$
$
\mathcal{G}^{(B)} = \left[
	\begin{matrix}
	0 & 1 & 0 \\
	1 & 0 & 1 \\
	0 & 1 & 0 
	\end{matrix}
  \right]~,
$
and
$
\mathcal{G}^{(C)} = \left[
	\begin{matrix}
	0 & 0 & 1 \\
	0 & 1 & 0 \\
	1 & 0 & 1 
	\end{matrix}
  \right]~.
$
\end{center}

To measure the diversity of interactions in a population we define the
\emph{interaction network complexity}
\begin{equation}
\Cmu ( {\mathcal G} ) = - \sum_{f_i, f_j, f_k > 0}
\frac{v_{ij}^k}{V^k} \log_2 \frac{v_{ij}^k}{V^k} ~,
\end{equation}
where
\begin{equation}
v_{ij}^k = \left\{
  \begin{array}{ll}
  f_i f_j , & T_k = T_j \circ T_i \mathrm{~has~occurred,}\\
  0, & \mathrm{otherwise} ~,
  \end{array}
  \right.
\end{equation}
$V^k = \sum_{lm} v_{lm}^k$ is a normalizing factor, and $f_i$ is the fraction
of \eM\ type $i$ in the soup. In order to emphasize our interest in actual
reproduction pathways, we consider only those that have occurred in the soup.

\subsection{Meta-Machines}

Given a population $\mathcal{P}$ of \eMs, we define a \emph{meta-machine}
$\Omega \subset \mathcal{P}$ to be a connected set of \eMs\ that is invariant
under composition. That is,
$\Omega$ is a meta-machine if and only if (i) $T_j \circ T_i \in \Omega$
for all $T_i,T_j \in \Omega$, (ii) for all $T_k \in \Omega$, there exists
$T_i,T_j \in \Omega$ such that $T_k=T_j \circ T_i$, and (iii) there is a
nondirected path between every pair of nodes in $\Omega$'s interaction
network $\mathcal{G}_\Omega$. The interactions in Fig. \ref{ExNet} describe
a meta-machine of Fig. \ref{ExTr}'s \eMs.

The meta-machine captures the notion of a self-replicating and
autonomous entity and is consistent with Maturana and Varela's
\emph{autopoietic set} \cite{VarelaMaturana1}, Eigen and Schuster's
\emph{hypercycle} \cite{Schuster1} and Fontana and Buss' \emph{organization}
\cite{Fontana96a}.

\begin{figure}
\begin{center}
\scalebox{0.55}{\includegraphics{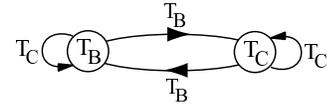}}
\caption{The meta-machine to which that in Fig. \ref{ExNet} decays
  under the population dynamics of Eq. (\ref{PThEq}).
  }
\label{DecayMetaMachine}
\end{center}
\end{figure}

\section{Population Dynamics}

We employ a continuously stirred flow reactor with an influx rate $\Phi_{in}$
that consists of a population $\mathcal{P}$ of $N$ \eMs. The dynamics of the
population is iteratively ruled by compositions and replacements as follows:
\begin{enumerate}
\setlength{\topsep}{-2mm}
\setlength{\itemsep}{-1mm}
\item \eM\ \emph{Generation}:
	\begin{enumerate}
		\setlength{\topsep}{-2mm}
		\setlength{\itemsep}{-1mm}
		\item With probability $\Phi_{in}$ generate a random \eM\ $T_R$
			(\emph{influx}).
		\item With probability $1-\Phi_{in}$ (\emph{reaction}):
			\begin{enumerate}
			\setlength{\topsep}{-2mm}
			\setlength{\itemsep}{-1mm}
				\item Select $T_A$ and $T_B$ randomly.
				\item Form the composition $T_C = T_B \circ T_A$.
			\end{enumerate}
		\end{enumerate}
\item \eM\ \emph{Outflux}:
	\begin{enumerate}
		\setlength{\topsep}{-2mm}
		\setlength{\itemsep}{-1mm}
		\item Select an \eM\ $T_D$ randomly from the population.
		\item Replace $T_D$ with either $T_C$ or $T_R$.
	\end{enumerate}
\end{enumerate}

Below, $T_R$ will be uniformly sampled from the set of all two-state \eMs.
This set is also used when initializing the population. The insertion of
$T_R$ corresponds to the influx while the removal of $T_D$ corresponds to
the outflux. The latter keeps the population size constant. Note that there
is no spatial dependence in this model; \eMs\ are picked uniformly from
the population for each replication. The finitary process soup
here is a well stirred gas of reacting objects.

When there is no influx ($\Phi_{in}$=0) and the population is closed with
respect to composition, the population dynamics is described by a
finite-dimensional set of equations:
\begin{equation}
\mathbf{f}_t^{(k)} =
  \mathbf{f}_{t-1} \cdot \mathcal{G}_{ij}^{(k)} \cdot \mathbf{f}_{t-1}^T Z^{-1},
\label{PThEq}
\end{equation}
where $\mathbf{f}_t^{(k)}$ is the frequency of \eM\ type $k$ at time $t$ and
$Z^{-1}$ is a normalization factor.

In addition to capturing the notion of self-replicating entities, meta-machines
also describe an important type of invariant set of the population dynamics.
Formally, we have
\begin{equation}
\Omega = \mathcal{G} \circ \Omega ~.
\end{equation}
These invariant sets can be stable or unstable under the population dynamics.
Note that the meta-machine of Fig. \ref{ExNet} is unstable: only $T_A$ produces
$T_A$s. As such, over time the population dynamics will decay to the
meta-machine of Fig. \ref{DecayMetaMachine}, which describes a soup consisting
only of $T_B$s and $T_C$s. This example also happens to illustrate that
copying---implemented here by the identity object $T_A$---need not dominate
the population and so does not have to be removed by hand, as done in several
prior pre-biotic models. It can decay away due to the intrinsic population
dynamics.

\begin{figure}
\begin{center}
\scalebox{0.325}{\includegraphics{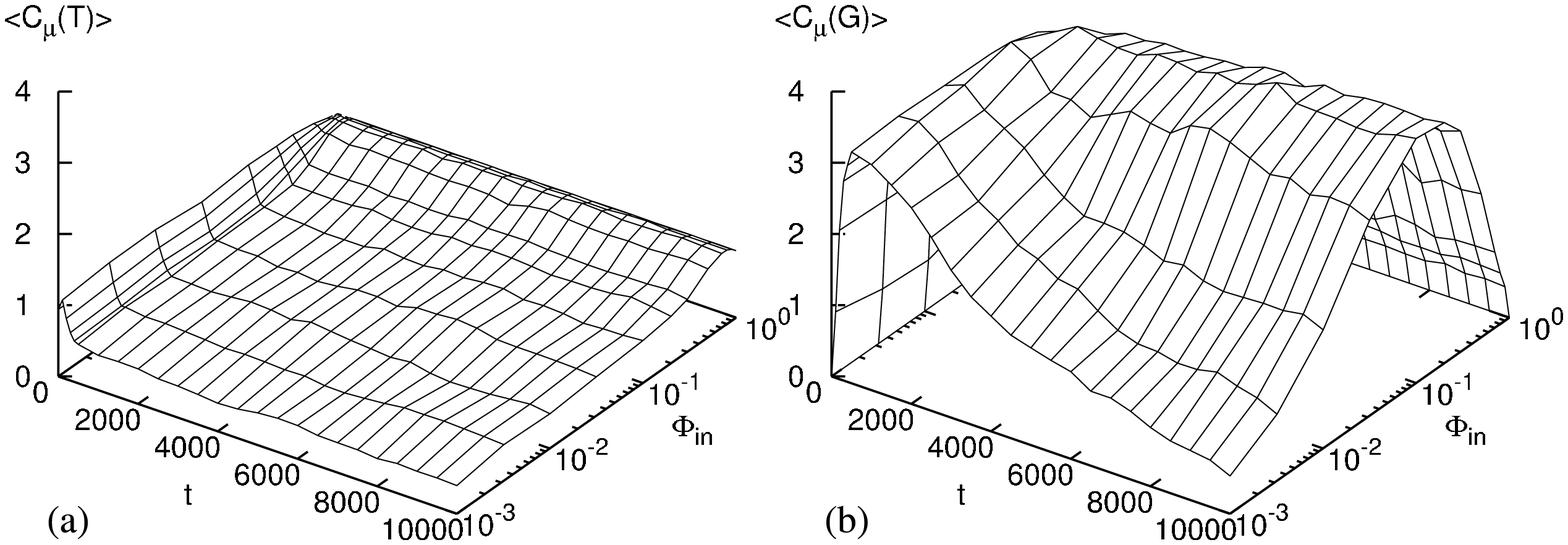}}\\
\caption{(a) Population-averaged \eM\ complexity $\langle C_{\mu}(T) \rangle$
  and (b) run-averaged interaction network complexity
  $\langle C_{\mu}(\mathcal{G}) \rangle$ as a function of time $t$ and
  influx rate $\Phi_{in}$ for a population of $N = 100$ objects.
  (Reprinted with permission from \protect\cite{Crut04a}.)
  }
\label{Surfs}
\end{center}
\end{figure}

\section{Simulations}

A system constrained by closure forms one useful base case that allows for
a straightforward analysis of the population dynamics. It does not permit,
however, for the innovation of structural novelties in the soup on either
the level of individual objects (\eMs) or on the level of their interactions.
What we are interested in is the possibility of open-ended evolution of
\eMs\ and their meta-machines. When enabled as an open system,
both with respect to composition and influx, the soup constitutes a
constructive dynamical system and the population dynamics of Eq. (\ref{PThEq})
do not strictly apply. (The open-ended population dynamics of epochal
evolution is required \cite{Crut99a}.)

\begin{figure}
\begin{center}
\scalebox{0.45}{\includegraphics{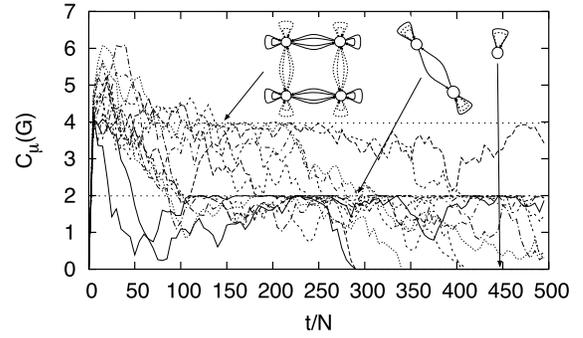}}
\caption{Meta-machine decomposition in a closed soup: 15 separate runs with
  $N = 500$. While the minimal 4-element meta-machine $\Omega_4$ (shown)
  dominates the soup, $C_{\mu}(\mathcal{G})$ is bounded by 4 bits. Once
  outflux removes one of its \eMs, rapidly $\Omega_4$ decays to $\Omega_2$,
  a 2-element meta-machine (shown). ($\Omega_4$ does not contain a
  sub-meta-machine of $3$ \eMs.) At this point $C_{\mu}(\mathcal{G})$ is
  bounded by $2$ bits. After some period of time, $\Omega_2$ decays to
  $\Omega_1$, a single self-reproducing \eM\ (shown), and
  $C_{\mu}(\mathcal{G})$ is fixed at 0.
  }
\label{HESuper}
\end{center}
\end{figure}

We first set the influx rate to zero in order to study dynamics that is
ruled only by compositional transformations. One important first observation
is that almost the complete set of machine types that are represented in the
soup's initial random population is replaced over time. Thus, even at the
earliest times, the soup generates genuine novelty. The population-averaged
individual complexity $\langle C_{\mu}(T)\rangle$ increases initially, as
Fig. \ref{Surfs}(a) ($\Phi_{\mathrm{in}} \approx 0$) from \cite{Crut04a}
shows. The \eMs\ are to some extent
shaped by the selective pressure coming from outflux and by geometric growth
due to composition. The turn-over is due to the dominance of nonreproducing
\eMs\ in the initial population. $\langle C_{\mu}(T)\rangle$ subsequently
declines since it is favorable to be simple as it takes a more extensive
stochastic search to find reproductive interactions that include more
complex \eMs.

Note (Fig. \ref{Surfs}(b), $\Phi_{\mathrm{in}} \approx 0$)
that the run-averaged interaction complexity
$\langle C_{\mu}(\mathcal{G})\rangle$ reaches a significantly higher value
than $\langle C_{\mu}(T)\rangle$, implying that the population's structural
complexity derives from its network of interactions rather than the complexity
of its constituent individuals. $\langle C_{\mu}(\mathcal{G})\rangle$ continues
to grow while compositional paths are discovered and created. A maximum is
eventually reached after which $\langle C_{\mu}(\mathcal{G})\rangle$ declines
and settles down to zero when one single type of self-reproducing \eM\ takes
over the whole population.

By monitoring the individual run values of $C_{\mu}(\mathcal{G})$ rather than
the ensemble average, one sees that they form plateaus as shown in Fig.
\ref{HESuper}. The plateaus---at $C_{\mu}(\mathcal{G}) = 4$ bits and,
most notably, at $C_{\mu}(\mathcal{G}) = 2$ bits and 0 bits---are determined
by the largest meta-machine that is present at a given time. Being a closed
set, the meta-machine does not allow any novel \eMs\ to survive and this gives
the upper bound on $C_{\mu}(\mathcal{G})$. As one \eM\ type is removed from
$\Omega$ by the outflux, the meta-machine decomposes and the upper bound on
$C_{\mu}(\mathcal{G})$ lowers. This produces a stepwise and irreversible
succession of meta-machine decompositions.

\begin{figure}
\begin{center}
\scalebox{0.3}{\includegraphics{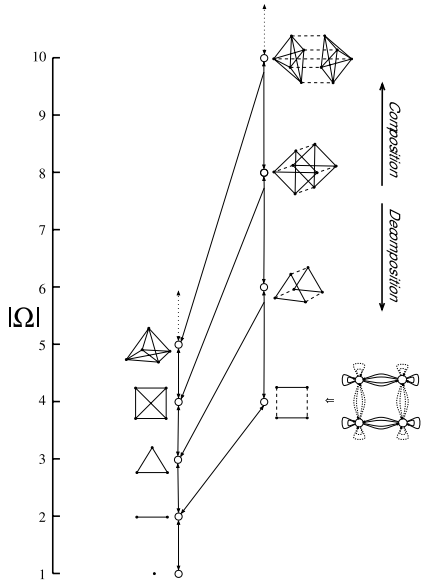}}
\caption{Meta-machine hierarchy of dynamical composition and decomposition.
  Dots denote \eMs. An isolated dot denotes a self-replicating \eM.
  Solid lines denote $T_A \stackrel{T_B}{\longrightarrow} T_C$ transitions.
  Dashed lines denote $T_B \stackrel{T_A}{\longrightarrow} T_C$ transitions.
  Although all possible transitions are used by the meta-machines
  shown, they are represented in a simplified way according to
  $\Omega_4$; cf. Fig. \ref{HESuper}.
  }
\label{MMHierarchy}
\end{center}
\end{figure}

Thus, in the case of zero influx, one sees that the soup moves from one extreme
to another. It is completely disordered initially, generates structural
complexity in its individuals and in its interaction network, runs out of
resources (poorly reproducing \eMs\ that are consumed by outflux), and
decomposes down to a single type of simple self-reproducing \eM.

Although Fig. \ref{HESuper} shows only three plateaus, there is in principle
one plateau for every meta-machine that at some point is the largest one
generated by the soup. The diagram in Fig. \ref{MMHierarchy} summarizes our
results from a more extensive and systematic survey of meta-machine
hierarchies from a series of runs with $N=500$. It gives one illuminating
example of how the soup spontaneously generates hierarchies of meta-machines.

Leaving closed soups behind, we now investigate the effects of influx. Recall
the population-averaged \eM\ complexity $\langle C_{\mu}(T) \rangle$ and the
run-averaged interaction network complexity
$\langle C_{\mu}(\mathcal{G}) \rangle$ as a function of $t$ and
$\Phi_{in}$ shown in Fig. \ref{Surfs}. Over time, $\langle C_{\mu}(T) \rangle$
behaves similarly for $\Phi_{in}>0$ as it does when $\Phi_{in}=0$. It
increases rapidly initially, reaches a peak, and declines to a steady state.
Notably, the emergence of complex organizations of interaction networks occurs
where the average structural complexity of the \eMs\ is low. Stationary
$\langle C_{\mu}(T) \rangle$ is instead maximized at a relatively high influx
rate ($\Phi_{in} \approx 0.75$) at which $\langle C_{\mu}(\mathcal{G}) \rangle$
is small compared to its maximum. As $\Phi_{in}$ is increased, so is
$\langle C_{\mu}(\mathcal{G}) \rangle$ at large times.
$\langle C_{\mu}(\mathcal{G}) \rangle$ is maximized around
$\Phi_{in} \approx 0.1$. For higher influx rates, individual novelty has a
deleterious effect on the sophistication of a population's interaction
network. Existing reproductive paths do not persist due to the low rate of
successful compositions of highly structured (and so specialized) individuals.
We found that the maximum network complexity $\widehat{C_{\mu}}(\mathcal{G})$
grows slowly and linearly over time at $\approx 7.6 \cdot 10^{-4}$
bits/replication.

\section{Summary and Conclusions}

To understand the basic mechanisms driving the evolutionary emergence of
structural complexity in a quantitative and tractable pre-biotic setting,
we investigated a well stirred soup of \eMs\ (finite-memory communication
channels) that react with each other by
composition and so generate new \eMs. When the soup is open with respect to
composition and influx, it spontaneously builds structural complexity on the
level of transformative relations among the \eMs\ rather than in the
\eM\ individuals themselves. This growth is facilitated by the use of
relatively non-complex individuals that represent general and elementary local
functions rather than highly specialized individuals. The soup thus
maintains local simplicity and generality in order to build up hierarchical
structures that support global complexity. Novel computational representations
are intrinsically introduced in the form of meta-machines that, in turn, are
interrelated in a hierarchy of composition and decomposition. Computationally
powerful local representations are thus not necessary (nor effective) in order
for the emergence and growth of complex replicative processes in the finitary
process soup. Meta-machines in closed soups eventually decay. For
$C_{\mu}(\mathcal{G})$ to maintain and grow the soup must be fed with novel
material in the form of random \eMs. Otherwise, any spontaneously generated
meta-machines are decomposed (due to finite-population sampling) and the
population eventually consists of a
single type of trivially self-reproducing \eM. At an intermediate influx
rate, however, the interaction network complexity is not only maintained
but grows linearly with time. This, then, suggests the possibility of
open-ended evolution of increasingly sophisticated organizations.

%
%
%
%

\section{Acknowledgments}

This work was supported at the Santa Fe Institute under the Network Dynamics
Program funded by the Intel Corporation and under the Computation, Dynamics,
and Inference Program via SFI's core grants from the National Science and
MacArthur Foundations. Direct support was provided by NSF grants DMR9820816
and PHY9910217 and DARPA Agreement F306020020583. OG was partially funded by
PACE (Programmable Artificial Cell Evolution), a European Integrated Project
in the EU FP6-IST-FET Complex Systems Initiative.


\bibliographystyle{alife10}
\bibliography{refs}

\end{document}